\documentclass[aps,prl,twocolumn,superscriptaddress,longbibliography]{revtex4-1}
\usepackage{braket}
\usepackage{graphicx}
\usepackage{amsmath}    
\usepackage{amsfonts}
\usepackage{lipsum}
\usepackage{xcolor}
\usepackage{qcircuit}

\begin{document}

\title{Fast-forwarding quantum simulation  \\ 
with real-time quantum Krylov subspace algorithms}

\author{Cristian L. Cortes}
\affiliation{Center for Nanoscale Materials,\\ Argonne National Laboratory, Lemont, Illinois 60439, USA}
\author{A. Eugene DePrince, III}
\affiliation{Department of Chemistry and Biochemistry,\\ 
Florida State University, Tallahassee, Florida
32306-4390, United States}
\author{Stephen K. Gray }
\affiliation{Center for Nanoscale Materials,\\ Argonne National Laboratory, Lemont, Illinois 60439, USA}

\begin{abstract}
Quantum subspace diagonalization (QSD) algorithms have emerged as a competitive family of algorithms that avoid many of the optimization pitfalls associated with parameterized quantum circuit algorithms. While the vast majority of the QSD algorithms have focused on solving the eigenpair problem for ground, excited-state, and thermal observable estimation, there has been a lot less work in considering QSD algorithms for the problem of quantum dynamical
simulation. In this work, we propose several quantum Krylov fast-forwarding (QKFF) algorithms capable of predicting long-time dynamics well beyond the coherence time of current quantum hardware. Our algorithms use real-time evolved Krylov basis states prepared on the quantum computer and a multi-reference subspace method to ensure convergence towards high-fidelity, long-time dynamics.  
In particular, we show that the proposed multi-reference methodology provides a systematic way of trading off circuit depth with classical post-processing complexity. We also demonstrate the efficacy of our approach through numerical implementations for several quantum chemistry 
problems including the calculation of the auto-correlation and dipole moment correlation functions.
\end{abstract}

\maketitle

\section*{Introduction}
Quantum simulation remains one of the most promising applications of quantum computation due its potential impact on high-energy physics, cosmology, condensed matter physics, atomic physics, and quantum chemistry. While the vast majority of quantum-simulation-based algorithms have been designed for the fault-tolerant quantum computing era \cite{berry2015simulating,low2017optimal,low2018hamiltonian,berry2019qubitization,low2019hamiltonian,lee2021even}, the current generation of noisy intermediate scale quantum (NISQ)  \cite{preskill2018quantum} computers limit the types of algorithms that could be implemented in the near term \cite{bauer2020quantum,cerezo2021variational,bharti2022noisy,motta2021emerging}. Variational quantum algorithms (VQAs) with parameterized quantum circuits have emerged as one of the leading methodologies capable of dealing with these constraints, and within this context, several NISQ-friendly quantum simulation algorithms have been proposed. These include the subspace variational quantum simulator 
\cite{heya2019subspace}, iterative approaches \cite{yuan2019theory,endo2020variational,otten2019noise}, and fast-forwarding approaches such as variational fast-forwarding 
\cite{cirstoiu2020variational}, variational Hamiltonian diagonalization 
\cite{commeau2020variational}, and fixed-state variational fast-forwarding 
\cite{gibbs2021long}. The overarching idea in all of these methods consists of using a variational wavefunction, $\ket{\psi(\boldsymbol{\theta})}=U(\boldsymbol{\theta})\ket{\boldsymbol{0}}$,  defined with respect to a parameterized quantum circuit  $U(\boldsymbol{\theta})$ and using a quantum-classical computer feedback loop to solve the optimization problem. In recent years, however, it has been shown that a wide variety of optimization problems relevant to VQAs can display non-convexity and vanishing gradients which can lead to fundamental optimization challenges \cite{bittel2021training,mcclean2018barren,wang2020noise,cerezo2021cost,arrasmith2020effect}. In addition, these algorithms suffer from large measurement overheads which can lead to long run times \cite{cerezo2021variational,bharti2022noisy}. 

In this regard, quantum subspace diagonalization (QSD) methods have emerged as an alternative approach to the conventional parameterized quantum circuit methodology \cite{mcclean2017hybrid,huggins2020non, parrish2019quantum,stair2020multireference,cohn2021quantum,epperly2021theory}. QSD methods express the variational wavefunction as a linear combination of non-orthogonal quantum states that are independently prepared on a quantum computer. By design, this formulation solves a convex optimization problem that avoids the challenges associated with conventional VQAs and parameterized quantum circuits (i.e. NP-hardness and barren plateau phenomena \cite{mcclean2018barren,cerezo2021cost}). While the vast majority of QSD algorithms have been applied to ground \cite{stair2020multireference,huggins2020non,cohn2021quantum,klymko2021real}, excited-state \cite{parrish2019quantum,cortes2022quantum,guzman2022accessing}, and finite temperature \cite{motta2020determining} observable estimation, there has been a lot less work in applying QSD algorithms to the problem of quantum dynamical simulation.
We should note, though, in classical-computer-based quantum dynamics simulations
there is a long history in the use of QSD, including the pioneering work on
the iterative Lanczos method by Park and Light\cite{park1986unitary}.

To the best of our knowledge, the only work to consider the quantum dynamical simulation problem using QSD methods is due to Lim et al. \cite{lim2021fast}, where the authors proposed a quantum subspace diagonalization method where the non-orthogonal states are constructed from the set of cumulative $K$-moment states, $\mathbb{CS}_K = \mathbb{S}_0 \cup \mathbb{S}_1 \cup \mathbb{S}_2 \cdots \cup \mathbb{S}_K$, where $\mathbb{S}_p = \{U_{i_p}\cdots U_{i_2}U_{i_1}\!\ket{\phi_o} \}$. Here, it is assumed that the Hamiltonian can be written as a sum of unitaries $U_i$ and $\ket{\phi_o}$ can be prepared efficiently on a quantum computer. If the set of unitaries $U_i$ are tensor products of Pauli operators, the problem of finding the subspace matrices reduces to a measurement of the quantum state $\ket{\phi_o}$ in different Pauli bases and, by construction, avoids the use of a Hadamard test.

\begin{figure*}
    \centering
    \includegraphics[width=17cm]{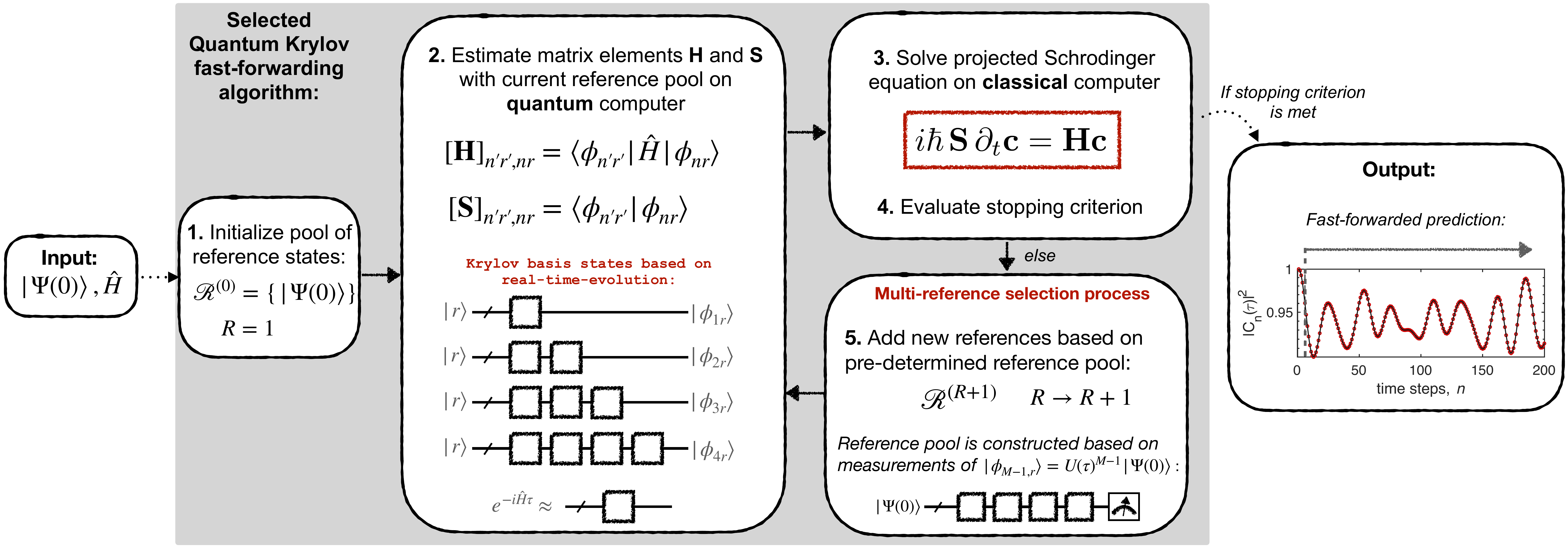}
    \caption{Overview of selected quantum Krylov fast-forwarding (sQKFF) method. }
    \label{fig:my_label}
\end{figure*}

In this manuscript, we build off the work by Lim et al. \cite{lim2021fast} and Stair et al. \cite{stair2020multireference} and propose a multi-reference quantum Krylov fast-forwarding (QKFF) algorithm for quantum dynamical simulation. In particular, we show that the addition of multi-reference states provides a route towards high-fidelity, long-time quantum simulation with a small number of Trotter steps. Our approach provides a controlled trade-off between quantum complexity, defined with respect to circuit depth, and classical complexity, defined with respect to the post-processing complexity (i.e. solving a large system of equations on a classical computer). Combined with the multi-fidelity estimation (MFE) protocol proposed in our previous work \cite{cortes2022quantum}, our approach avoids the Hadamard test with the added benefit of using an ultra-compact wavefunction representation \cite{klymko2021real} that is not classically tractable.  To demonstrate the potential of our approach, we present numerical experiments for various physical problems including the calculation of the auto-correlation function and two-time dipole moment correlation function. 

\section*{Quantum Krylov Fast Forwarding}
Quantum dynamical simulation aims to solve the time-dependent Schrodinger equation ($\hbar$ = 1 throughout),
\begin{equation}
    i~  \partial_t \ket{\psi(t)} = \hat{H} \ket{\psi(t)},
    \label{Schrodinger}
\end{equation}
which describes the dynamics of a general many-body Hamiltonian $\hat{H}$ written as a sum of $N$-qubit Pauli terms, $\hat{H} = \sum_i^L h_i \hat{P}_i$, where $h_i$ is a weighting coefficient and $\hat{P}_i$ is a general tensor product of $N$ Pauli operators, ${\hat{P}}_i = \otimes_{k=1}^{N_i} \hat{\sigma}_{i_k}^{(m_k)}$, with $m_k$ denoting the qubit number and $i_k$ acting as a label for the type of Pauli operator $\{\hat{I}, \hat{\sigma}_x,\hat{\sigma}_y,\hat{\sigma}_z\}$.  We assume that the Hamiltonian is time-independent but do not impose any type of restrictions on the locality of the Hamiltonian, thereby making this approach applicable to a wide variety of physical problems of interest. The multi-reference quantum Krylov method proceeds by approximating the wavefunction $\ket{\psi(t)}$ by a linear combination of non-orthogonal quantum states,
\begin{equation}
    \ket{\psi(t)} \approx \ket{\psi_{K}(t)} = \sum_{n=0}^{M-1}\sum_{r=1}^R c_{nr}(t) \ket{\phi_{nr}},
    \label{LinearExpansion}
\end{equation}
where $M$ corresponds to the single-reference Krylov subspace dimension and $R$ corresponds to the total number of reference states. The choice of non-orthogonal states $\ket{\phi_{nr}}$ ultimately leads to a wide variety of hybrid quantum-classical algorithms with various algorithmic trade-offs \cite{cortes2022quantum}. In this work, we will consider the real-time evolved Krylov basis states which form an order-$M$ Krylov subspace, $\mathcal{K}_M = \text{span}\{\ket{r},e^{-i\hat{H}\tau}\ket{r},,e^{-i\hat{H}2\tau}\ket{r}\cdots,e^{-i\hat{H}(M-1)\tau}\ket{r} \}$, where $\ket{r}$ corresponds to the $r$th reference state, and the $(n,r)$ non-orthogonal state is given by, $\ket{\phi_{nr}} = e^{-i\hat{H}n\tau}\ket{r}$.

Substituting Eq.(\ref{LinearExpansion}) into Eq. (\ref{Schrodinger}) and multiplying from the left by $\bra{\phi_{n'r'}}$, we obtain the quantum subspace Schrodinger equation,
\begin{equation}
    i~ \mathbf{S} \partial_t{\mathbf{c}}(t) = \mathbf{H}\mathbf{c}(t),
    \label{SubspaceSchrodinger}
\end{equation}
where $\mathbf{c}(t)$ is a $RM\times 1$ column vector of time-dependent expansion coefficients, while $\mathbf{H}$ and $\mathbf{S}$ are $RM\times RM$ subspace matrices defined as, $[\mathbf{H}]_{n'r',nr} = \braket{\phi_{n'r'}|\hat{H}|\phi_{nr}}$ and $[\mathbf{S}]_{n'r',nr} = \braket{\phi_{n'r'}|\phi_{nr}}$ respectively. Fast-forwarding is achieved by solving the quantum subspace Schrodinger equation, Eq. (\ref{SubspaceSchrodinger}), with respect to the expansion coefficients $\mathbf{c}(t)$. Numerically, the solution can be obtained in a variety of different ways including the use of linear multi-step methods or Runge-Kutta methods \cite{butcher2016numerical}, which becomes relevant when the system size becomes large. However, to obtain a better theoretical understanding, we focus on the formal solution written succinctly as,
\begin{equation}
    \mathbf{c}(t) = e^{-i\mathbf{S}^{-1}\mathbf{H}t}\mathbf{c}(0),
    \label{solution}
\end{equation}
where the initial condition column vector $\mathbf{c}(0)$ is given by, $\mathbf{c}(0) = \mathbf{S}^{-1}\mathbf{d}(0)$, and $\mathbf{d}(0) = (\braket{\phi_{01}|\psi(0)},\braket{\phi_{11}|\psi(0)},\cdots,\braket{\phi_{(M-1)R}|\psi(0)})^T$. Note that $\mathbf{d}(0)$ corresponds to the first column of the overlap matrix $\mathbf{S}$ if the first reference state is equal to the initial state, $\ket{\psi(0)}$. 
The evaluation of the matrix exponential in (\ref{solution}) provides an estimate of the complex-valued expansion coefficients $\mathbf{c}(t)$ for {arbitrary} times $t$, allowing for fast-forwarded predictions well beyond the coherence time of the quantum hardware. It is worth noting that the overlap matrix $\mathbf{S}$ can be poor conditioned, as demonstrated in previous works \cite{stair2020multireference,cortes2022quantum}, requiring special considerations when taking the 
matrix inverse.
As suggested by Klymko et al. \cite{klymko2021real},
we use singular value decomposition \cite{trefethen1997,businger1969} of
$\mathbf{S}$, coupled with zeroing out singular values that fall
below a threshold when the inverse is constructed (see also Appendix A).

While a single-reference-based quantum Krylov fast-forwarding (QKFF) algorithm could be used in practice, we found that such an approach is severely limited in predicting long-time dynamics with a small number of Trotter steps. As a result, our main contribution corresponds the proposal of a \emph{multi-reference, selected} quantum Krylov fast-forwarding (sQKFF) algorithm for predicting long-time quantum dynamics with high fidelity, as summarized in Figure 1. The algorithm starts with a single reference state which is equal to the initial quantum state supplied by the user. The projected subspace matrix elements ($[\mathbf{H}]_{n'r',nr}$ and $[\mathbf{S}]_{n'r',nr}$) are then estimated on the quantum computer using either the Hadamard test or MFE protocol \cite{cortes2022quantum}. A classical computer is then used to find the solution to the quantum subspace Schrodinger equation. If a stopping criterion is met, the sQKFF algorithm terminates, otherwise the algorithm continues by selecting additional reference states based on a selection process that we will discuss later. Once the new reference states are selected, the previous three steps are repeated until the stopping criterion is met.

\paragraph{Reference Selection Process.}
The reference selection process is critical for the sQKFF algorithm. While many choices exist, we propose a simple approach that requires negligible circuit depth when compared to the Trotterized circuits required for real-time Hamiltonian simulation. Here, the basic idea corresponds to performing transition probability measurements based on the user-defined initial state, $\ket{\psi(0)}$. The transition probability measurement procedure consists of two sub-steps: (1) preparing the $M$th Krylov subspace state, $\ket{\phi_{M-1,1}} = e^{-i\hat{H}(M-1)\tau}\ket{\psi(0)}$, on the quantum computer, and (2) performing sampling measurements in the Pauli-Z basis (see Figure 1). The frequency of the measured bitstrings will follow an underlying transition probability distribution, $p(x) = |\braket{x|e^{-i\hat{H}(M-1)\tau}|\psi(0)}|^2$. Assuming $K$ total samples, the bitstrings with the largest observed transition probabilities, $n_x/K$, are added to the reference pool $\mathcal{R}^{(R)}$, where $n_x$ refers to the number of times the $x$th bitstring was observed. 
While a complete determination of $p(x)$ scales exponentially with system size, we
have found that relatively modest sampling suffices, typically $K$ on the order of hundreds or  thousands of samples provided good estimates.
Individual bitstrings are added as reference states iteratively until our stopping criterion is met. To ensure that the superposition of bitstrings preserve any symmetries inherent in the underlying Hamiltonian, a separate subspace Hamiltonian can be constructed with the sampled bitstrings which is then diagonalized on the classical computer. The resulting eigenvectors will provide the numerical values for the amplitudes of the bitstrings, ensuring that all of the Hamiltonian symmetries are preserved.

\paragraph{Stopping Criterion.} A pragmatic stopping condition is to specify some
maximum time, $T_{max}$, of physical interest and an acceptable tolerance, $\epsilon$, for
some desired dynamical property  such as the magnitude of a correlation function. If for
times $t \leq T_{max}$ addition of more reference states yields changes no greater
than $\epsilon$ in the dynamical property then stop.

\section*{Time-dependent Observables}
The solution to the projected subspace Schrodinger equation provides an estimate of the complex-valued coefficients $\mathbf{c}(t)$. Once these coefficients have been determined, it is then possible to calculate a wide variety of time-dependent observables through additional post-processing that may or may not require additional calls to the quantum computer. In the following, we consider three different time-dependent quantities that are relevant to quantum chemistry, nuclear physics, and materials science calculations: (1) the auto-correlation function, (2) time-dependent local and global observables, and (3) two-time correlation functions. 

1. The auto-correlation function, $C(t) = \braket{\psi(0)|\psi(t)}$, is perhaps the only time-dependent quantity that does not require any additional calls to the quantum computer based on the quantum Krylov method that we have outlined above. In general, we can write the approximate auto-correlation function as, $C(t) = \sum_n c_n(t) \braket{\psi(0)|\phi_n} = \mathbf{d}^\dagger(0)\cdot\mathbf{c}(t)$, which is clearly expressed in terms of quantities that originate from the sQKFF algorithm. 

2. Time-dependent observables are often desirable for predicting physical quantities such as charge densities and order parameters. Based on the linear combination of non-orthogonal states expression from Eq. (2), a general time-dependent observable may be written as: 
\begin{align}
    O(t) &= \braket{\psi(t)|\hat{O}|\psi(t)} \nonumber \\
    &= \sum_{k',k} c^*_{k'}(t)c_{k}(t)\braket{\phi_{k'}|\hat{O}|\phi_{k}}
    \label{observable}
\end{align}
where we used the single index $k=nr$ to simplify the notation. Based on Eq. (\ref{observable}), it is clear that the fast-forwarded prediction for $O(t)$ might require additional calls to the quantum computer for evaluating the matrix elements, $[\mathbf{O}]_{n'r',nr} = \braket{\phi_{n'r'}|\hat{O}|\phi_{nr}}$. Assuming that the observable $\hat{O}$ is expressed as a linear combination of Pauli words, $\hat{O} = \sum_i^{L_o} o_i \hat{P}_i$, this implies that additional calls would be required for Pauli words that do not coincide with the Pauli words from the original Hamiltonian decomposition, $\hat{H} = \sum_i^L h_i \hat{P}_i$. In the worst case, this would require an additional $\mathcal{O}(L_o (RM)^2)$ calls to the quantum computer, where $L_o$ is equal to the total number of Pauli terms for observable, $\hat{O}$.
\begin{figure*}
    \centering
    \includegraphics[width=17cm]{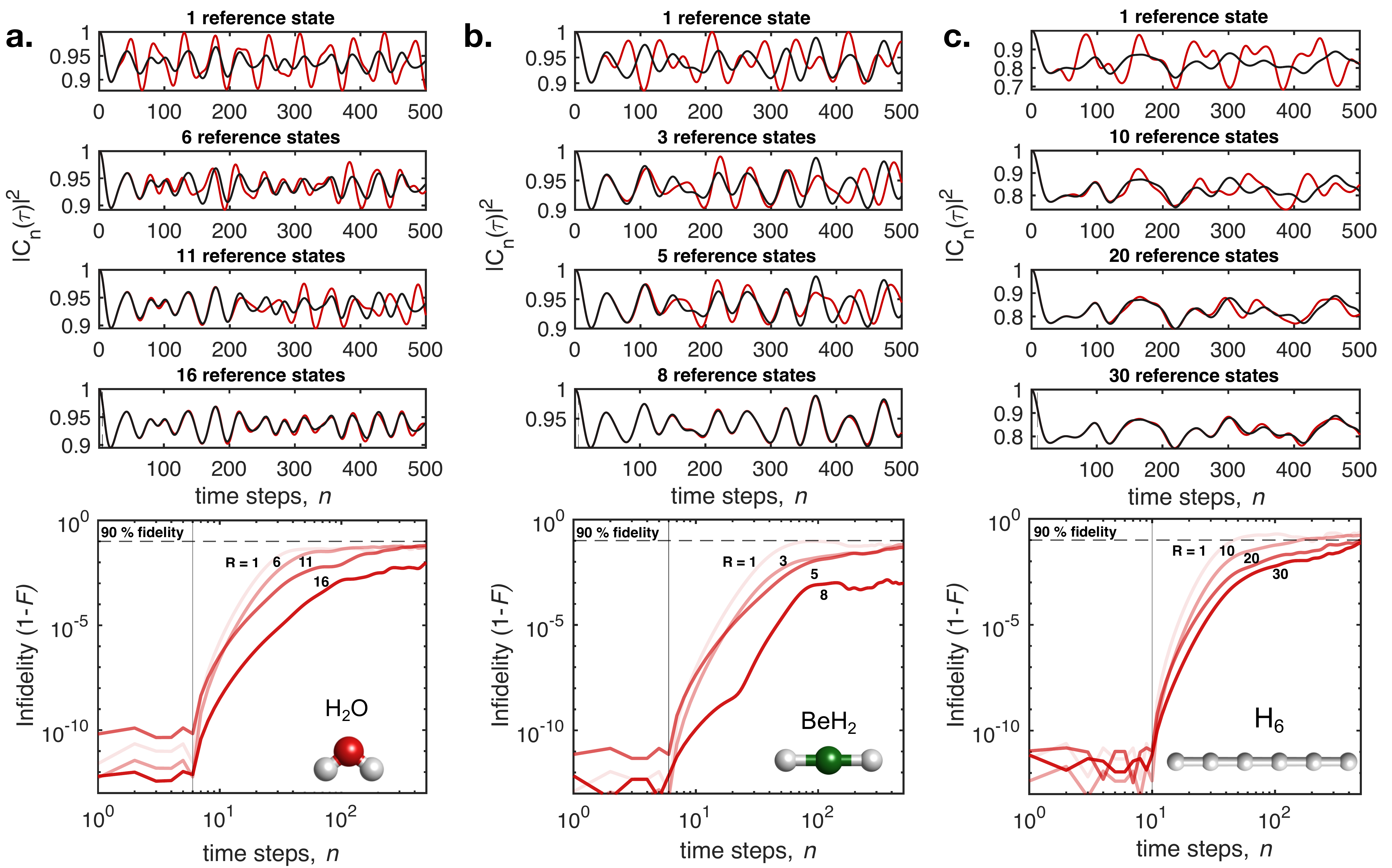}
    \caption{Fast-forwarding of the auto-correlation function, $C_n(\tau)=\braket{\psi(0)|\psi(n\tau)}$, for (a) H$_2$O molecule, (b) BeH$_2$ molecule, and (c) H$_6$ hydrogen chain. Top four rows denote the absolute value squared of the auto-correlation function for various numbers of reference states. The bottom row display the infidelity of the true wavefunction $\ket{\psi(t)}$ with respect to the sQKFF predicted wavefunction, $\ket{\psi_K(t)}$. In all cases, a time step size of $\tau = 0.1$ atomic units was used. For (a) and (b), the fast-forwarded prediction used the Krylov subspace dimension, $M=6$, while (c) used maximum Krylov dimension, $M=10$.}
    \label{fig:my_label}
\end{figure*}

3. Lastly, we consider the evaluation of two-time correlation functions of the form, $\braket{A(t+\tau)B(t)}$, which are used for the calculation of two-particle correlation functions, Green's functions and dipole moment correlation functions relevant to many different types of spectroscopies \cite{meyer1975pno, hess2002biological, kassal2011simulating, rickayzen2013green, parrish2019quantum2,  jamet2022quantum}. A general two-time correlation function between observable $\hat{A}$ and $\hat{B}$ may be written as, 
\begin{align}
    \braket{A(t+\tau)B(t)} &= \braket{\psi(t+\tau)|\hat{A} e^{-i\hat{H}\tau}\hat{B}|\psi(t)} \label{two-time} \\
    &= \sum_{k',k} c^*_{k'}(t+\tau)c_{k}(t)\braket{\phi_{k'}|\hat{A}e^{-i\hat{H}\tau}\hat{B}|\phi_{k}} \nonumber.
\end{align}
While $c_k(t)$ and $c^*_{k'}(t+\tau)$ are readily obtained from the original sQKFF algorithm evaluated with respect to the initial condition $\ket{\psi(0)}$, it is clear from Eq. (\ref{two-time}) that the fast-forwarded prediction for  the two-time correlation function $\braket{A(t+\tau)B(t)}$ requires the evaluation of $\braket{\phi_{k'}|\hat{A}e^{-i\hat{H}\tau}\hat{B}|\phi_{k}}$, which in turn requires a fast-forwarded prediction of the time-evolved quantum state $\ket{\tilde{\psi}(\tau)} = e^{-i\hat{H}\tau}\hat{B}\ket{\phi_k} = \sum_{k''}c_{k''}(\tau)\ket{\phi_{k''}}$, leading to the evaluation of $\braket{\phi_{k'}|\hat{A}e^{-i\hat{H}\tau}\hat{B}|\phi_{k}} = \sum_{k''}c_{k''}(\tau) \braket{\phi_{k'}|\hat{A}|\phi_{k''}}$. This shows that an observable estimate of $\braket{\phi_{k'}|\hat{A}|\phi_{k''}}$ would also be required for the evaluation of the two-time correlation function when $\hat{A}\neq\hat{B}$. This implies that in the most general case, without additional simplifications, two separate runs of the sQKFF algorithm will be required for the prediction of two-time correlation functions. The first run will provide a fast-forwarded prediction for $\ket{\psi(t)} = e^{-i\hat{H}t}\ket{\psi(0)}$, while the second run will provide a fast-forwarded prediction of $\ket{\tilde{\psi}(\tau)}$. For many physical problems of interest, however, it will be possible to reduce this requirement.

To illustrate this point, we consider the calculation of the two-time dipole moment correlation function, $ \braket{\hat{\mu}_\xi(t+\tau)\hat{\mu}(t)} = \braket{\Psi_G|\hat{\mu}_\xi(t+\tau)\hat{\mu}_\xi(t)|\Psi_G}$, where $\ket{\Psi_G}$ corresponds to the ground-state wavefunction of an electronic structure Hamiltonian $\hat{H}$ and the dipole moment operator, $\hat{\mu}_\xi = \sum_{pq} \mu_{pq}^\xi a^\dagger_p a_q$, is defined in the fermionic second quantized spin-orbital basis (additional details of the dipole moment correlation function and its evaluation is discussed in Appendix B). The dipole moment correlation function is a fundamental quantity of interest for chemistry and materials science because of its relation to the linear absorption spectrum. Details of this relationship can be found in Appendix C. To compute the two-time dipole moment correlation function, we first require preparing the ground-state wavefunction on the quantum computer. The preparation of the ground state can proceed in several ways and may certainly represent a challenge of its own. Here, we outline two methods that are amenable to near-term quantum computing. The first method assumes the use of a separate quantum Krylov diagonalization algorithm for ground-state energy estimation \cite{parrish2019quantum,stair2020multireference,cortes2022quantum,klymko2021real}, which is able to express the ground-state wavefunction as a linear combination of non-orthogonal states, $\ket{\psi_G} \approx \sum_{n=0}^{M_G-1} c_n \ket{\phi_n}$, where the coefficients $c_n$ are notably different from those in Eq. (\ref{LinearExpansion}). Assuming that we have $M_G$ non-orthogonal states in order to represent the ground-state wavefunction, the total run-time of the sQKFF algorithm will increase by a multiplicative factor of $M_G^2$. The second methodology does not suffer from the increased run-time and relies on using the results of a variational quantum eigensolver algorithm to find the approximate ground-state wavefunction, $\ket{\psi_G} \approx \ket{\psi(\boldsymbol{\theta})} =  U(\boldsymbol{\theta})\ket{0}^{\otimes{N}}$, where $U(\boldsymbol{\theta})$ represents the parameterized quantum circuit unitary. Assuming that the approximate ground-state wavefunction from either method is written as $\ket{\tilde{\Psi}_G}$, it is then possible avoid the requirement of running two separate sQKFF algorithms by using the commonly used approximation, $\ket{\psi(t)} = e^{-i\hat{H}t}\ket{\tilde{\Psi}_G} \approx e^{-i\tilde{E}_Gt}\ket{\tilde{\Psi}_G}$, where $\tilde{E}_G$ is the ground-state energy that would have been estimated from either method. Once this approximation is invoked, the two-time dipole moment correlation function becomes,  $ \braket{\hat{\mu}_\xi(t+\tau)\hat{\mu}(t)} = e^{i\tilde{E}_G \tau}\braket{\tilde{\Psi}_G|\hat{\mu}_\xi e^{-i\hat{H}\tau}\hat{\mu}_\xi|\tilde{\Psi}_G}$. This line of reasoning shows that only a single sQKFF run will be required for the fast-forwarded prediction of $\ket{\tilde{\psi}(\tau)} = e^{-i\hat{H}\tau}\hat{\mu}_\xi\ket{\tilde{\Psi}_G}$.

\section{Numerical Experiments}
In Figure 2, we compare the single reference QKFF and multi-reference sQKFF algorithms for prototypical quantum chemistry Hamiltonians consisting of (a) H$_2$O molecule with fixed bond angle $\phi=104.45$, (b) a BeH$_2$ molecule, and (c) a linear Hydrogen chain. In all three cases, the bond length is chosen to be equal to 1.85 \AA. Details of the quantum chemistry Hamiltonian, basis sets and 
active space selection for these systems may be found in Appendix A of our previous work \cite{cortes2022quantum}. Here, we focus on the fast-forwarded prediction of the auto-correlation function, $C(t) = \braket{\psi(0)|\psi(t)}$, using numerical state vector simulations with ideal time-evolution circuits. Future work will provide a more detailed analysis of the Trotter error, shot noise error, and other hardware noise effects. For all of the numerical simulations, we chose the single-reference Krylov subspace dimension to be equal to $M=6$. 
\begin{figure}[b]
    \centering
    \includegraphics[width=8cm]{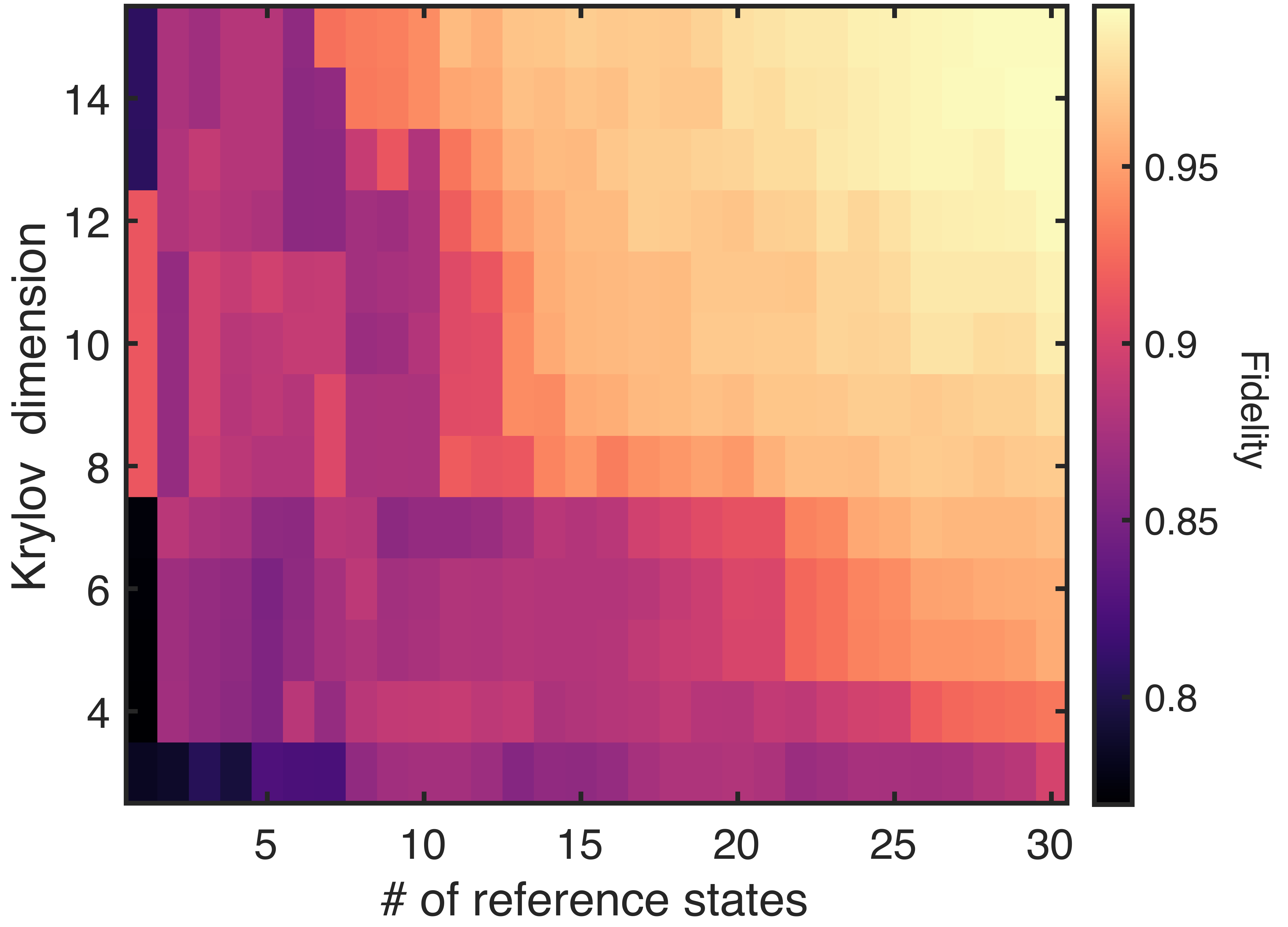}
    \caption{Trade-off between quantum and classical complexity.  
    Quantum complexity corresponds to circuit depth. Classical complexity is proportional to the number of reference states.}
    \label{Figure3}
\end{figure}
\begin{figure*}[t!]
    \centering
    \includegraphics[width=18cm]{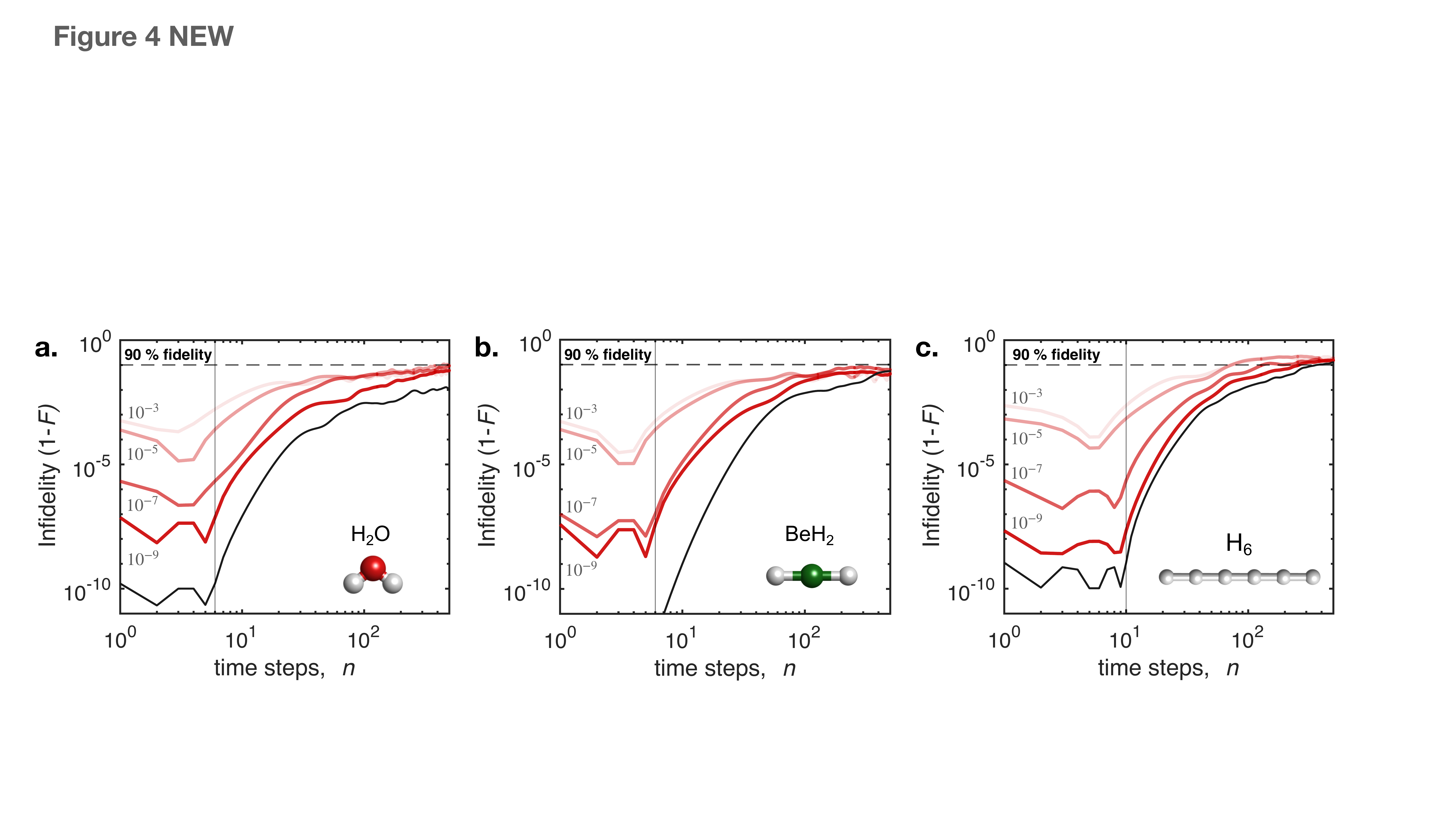}
    \caption{Noise robustness of sQKFF algorithm for same molecular systems as FIG 2. For each of the molecular systems, the calculations were performed using a total of: (a) 16, (b) 8, and (c) 30 reference states respectively with time step size $\tau = 0.1$ atomic units. Additive complex Gaussian noise with zero mean and standard deviation, $\sigma = \{10^{-3},10^{-5},10^{-7},10^{-9}\}$ is included for all of the matrix elements of $\mathbf{H}$ and $\mathbf{S}$ respectively. The singular value threshold of (a) $\epsilon = \{1.5, 1.5\times10^{-2}, 1.5\times10^{-4}, 1.5\times10^{-6}, 1.5\times10^{-8}\}$, (b) $\epsilon = \{10^{0}, 10^{-2}, 10^{-4}, 10^{-6}, 10^{-8}\}$, and (c) $\epsilon = \{10^{-1}, 10^{-3}, 10^{-5}, 10^{-7}, 10^{-9}\}$ was used respectively for each of the noise standard deviation parameters.}
    \label{Figure4}
\end{figure*}
Reference states were added with the selection process discussed above. The top four rows display the explicit time-evolved correlation function with a different number of reference states. The bottom row displays the infidelity of the true wavefunction with respect to the QKFF wavefunction, where the state fidelity is defined as, $F(t) = |\!\braket{\psi(t)|\psi_K(t)}\!|^2$. From the top row of Figure 2, it is  clear that the single-reference fast-forwarded prediction (red line) only matches the true correlation function (black line) for extremely short times. As additional reference states are added, we observe that the predicted auto-correlation function more closely aligns with the true correlation function. We emphasize that this increased fidelity does not require additional circuit depth, and only requires additional calls to the quantum computer to estimate the projected subspace matrix elements $[\mathbf{H}]_{n'r',nr}$ and $[\mathbf{S}]_{n'r',nr}$ respectively.

The route towards high-fidelity, long-time quantum simulation can be achieved in two distinct  ways: (1) increasing Krylov subspace dimension $M$, or (2) increasing the number of reference states, $R$. Ultimately, this leads to a trade-off between quantum and classical complexity for obtaining reliable high-fidelity, long-time predictions. In Figure 3, we highlight this trade-off by calculating the state fidelity for the water molecule as a function of Krylov dimension $M$ on the $y$-axis and the number of reference states $R$ on the $x$-axis. In all cases, we chose the singular value threshold, $\epsilon=1\times 10^{-9}$. In general, however, the singular value threshold should be optimized for different numbers of reference states and Krylov subspace dimension since it affects the fidelity prediction. These plots merely serve to provide a proof-of-concept of the quantum-classical complexity trade-off. From Figure 3, it is shown that high fidelities above $90\%$ can be achieved by increasing either of the two independent axes. In the near term where quantum hardware provides a severe limit on the gate depth, this plot illustrates how additional resources can be allocated to the classical computer in order to improve quantum dynamical simulation with higher accuracy. 

In addition to the quantum-classical trade-off shown in the previous plot, we also studied the noise robustness of the sQKFF algorithm. In Figure 4, we plot the infidelity of the H$_2$O, BeH$_2$ and H$_6$ molecules with the same parameters as Figure 2 (black lines). To study the effect of random noise, we included additive Gaussian noise with zero mean and standard deviation, $\sigma = \{10^{-3},10^{-5},10^{-7},10^{-9}\}$, shown in red. The magnitude of the standard deviation ultimately controls the digit precision of the projected subspace matrix elements $[\mathbf{H}]_{n'r',nr}$ and $[\mathbf{S}]_{n'r',nr}$. For all three molecular systems, we found that the fidelity could still remain close to $90\%$ at long times even with Gaussian noise as large as $\sigma = 10^{-3}$, which highlights the noise robustness of the sQKFF algorithm relevant for near-term quantum hardware.

\begin{figure*}
    \centering
    \includegraphics[width=13cm]{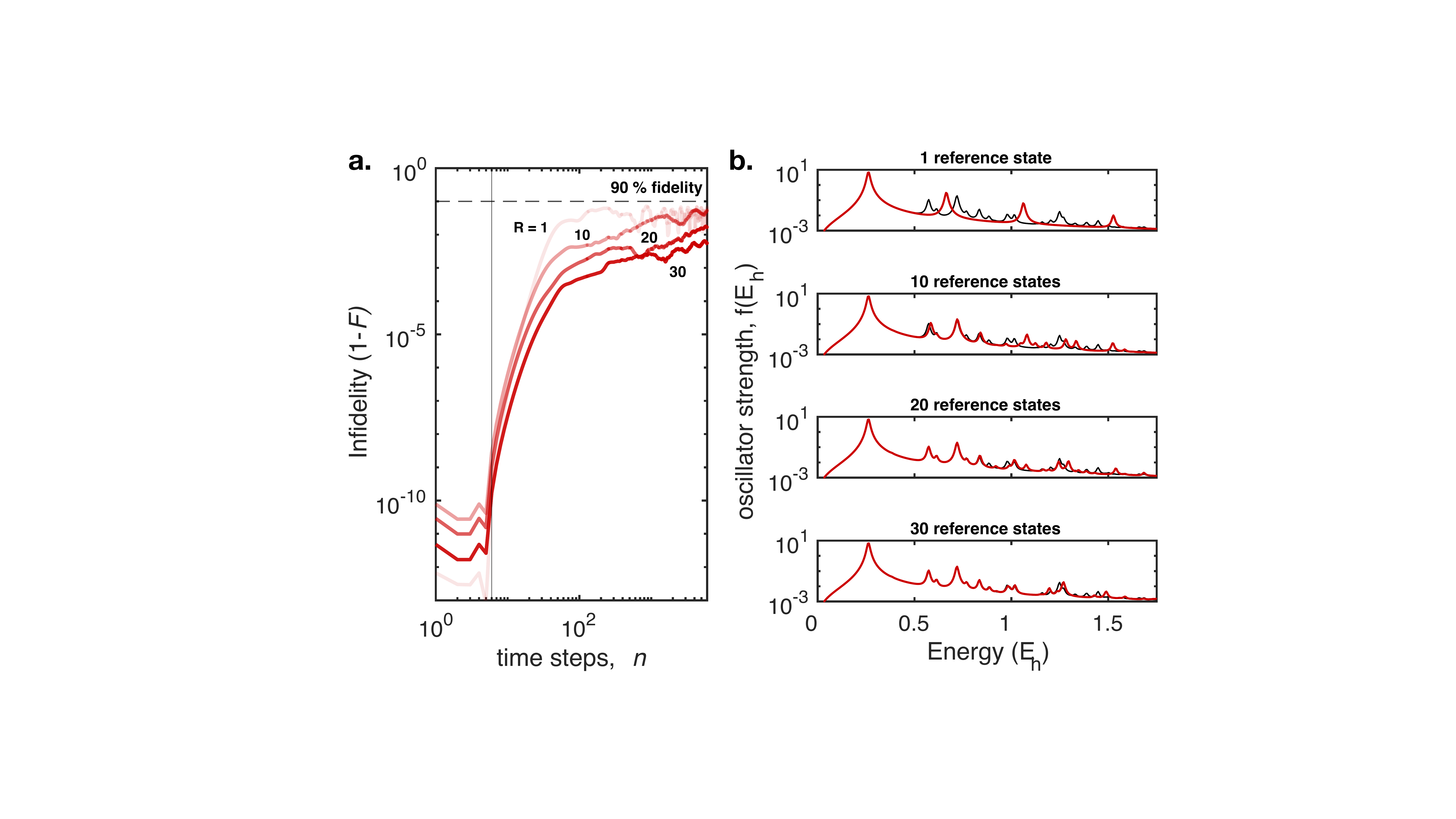}
    \caption{(a) Infidelity of dipole moment propagated wavefunction $\ket{\mu(t)} = e^{-i\hat{H}t}\hat{\mu}\ket{\Psi_G}$ compared to sQKFF prediction for a linear H$_6$ hydrogen chain. (b) Oscillator strength absorption spectrum with different numbers of reference states $R=\{1,11,21,31\}$. In all cases, a Krylov dimension size of $M=6$ was used along with the time step size, $\tau=0.1$ atomic units}
    \label{Figure5}
\end{figure*}

Finally, we test the sQKFF algorithm for the prediction of the two-time dipole moment correlation function, $ \braket{\hat{\mu}_\xi(t+\tau)\hat{\mu}(t)} = e^{i\tilde{E}_G \tau}\braket{\tilde{\Psi}_G|\hat{\mu}_\xi e^{-i\hat{H}\tau}\hat{\mu}_\xi|\tilde{\Psi}_G}$, where, as discussed previously, only a single run of the sQKFF algorithm is required. 
We emphasize that these plots serve as a proof of concept for this algorithm, however, based on the results from Appendix C, we would recommend using a standard quantum Krylov method for ground and excited-state energy estimation \cite{stair2020multireference,huggins2020non,cohn2021quantum,klymko2021real,parrish2019quantum,cortes2022quantum} or excited-state VQE methods \cite{nakanishi2019subspace,higgott2019variational,parrish2019quantum2} for the purpose of reconstructing the oscillator strength spectrum with current quantum hardware, especially when only low-lying excited state energies and oscillator strengths are desired. The advantage of this algorithm would correspond to its blackbox nature, where zero knowledge is required to estimate the ground or excited state energies for the physical system of interest. In Figure 5-a, we plot the infidelity of the dipole moment propagated wavefunction, $\ket{\mu(t)} = e^{-i\hat{H}t}\hat{\mu}\ket{\Psi_G}$, with respect to the quantum Krylov wavefunction for the linear H$_6$ hydrogen chain with the same parameters as Figure 2. In Figure 5-b, we plot the oscillator strength absorption spectrum which is defined as the Fourier transform of the time-evolved dipole moment correlation function (see Appendix C for details). To ensure that we obtain a stable and well-defined absorption spectrum, we added finite linewidth to the correlation function discussed in Eq. (9) of Appendix C which results in an additional $e^{-\gamma t}$ multiplicative factor where $\gamma$ corresponds to the finite linewidth. For all of the numerical experiments, we chose a linewidth equal to $\gamma = 1.5\times 10^{-2}$. Here, we observe the same general trends from the auto-correlation function prediction shown in Figure 2. While the single-reference QKFF algorithm provides an accurate prediction of the first transition peak (see top row of Figure 5-a), it is not able to accurately predict the higher energy transition peaks as highlighted by the black line. As more reference states are added, however, we find that the absorption spectrum more closely aligns with the true absorption spectrum with minor deviations at very high energies. Ultimately, this illustrates that the sQKFF algorithm works well for the prediction of a wide variety of time-dependent observables ranging from the auto-correlation function to the two-time correlation function relevant to a wide variety of physical applications.




\section*{Concluding Remarks}

To conclude, we have shown that real-time quantum Krylov subspace algorithms can be used to fast-forward quantum simulation well beyond the coherence time of current quantum hardware. We developed a theory quantum Krylov fast-forwarding and proposed a selected quantum Krylov fast-forwarding (sQKFF) algorithm that is capable of providing an estimate of time-dependent quantum states, which works especially well when the initial state contains a polynomial number of eigenstates \cite{gu2021fast,gibbs2021long}. We validated the algorithm with numerical experiments focusing on the calculation of the auto-correlation and dipole moment correlation functions for various molecules. While our work provides a way of studying a wide range of time dependent phenomena with near term quantum hardware, there remains many important avenues of research for improving the sQKFF algorithm. For instance, while we provided evidence of the noise robustness of the sQKFF algorithm, further work should elaborate on the effects of Trotter error, shot noise, and other realistic hardware imperfections. Furthermore, while our results indicate that adding reference states provides a way of achieving long-time, high-fidelity quantum dynamics simulation, the subspace dimension of the proposed algorithms required relatively large subspaces when compared to the to symmetry-projected subspaces inherent to the chemical systems. Even for these modest systems, the choice of individual bitstrings as reference states resulted in relatively slow convergence, and therefore alternative approaches may be required for larger systems. Further work should focus on providing a systematic study on the choice of the selection process and reference states in order to improve the sQKFF algorithm's performance. 
Additional work should also aim to develop and extend this algorithm so that it becomes applicable to time-dependent Hamiltonians. 

\section{ACKNOWLEDGMENTS}
\noindent This material is based upon work supported by Laboratory Directed Research and Development (LDRD) funding from Argonne National Laboratory, provided by the Director, Office of Science, of the U.S. Department of Energy under Contract No. DE-AC02-06CH11357 and the  U.S.~Department of Energy, Office of Science, Office of Advanced Scientific Computing Research and Office of Basic Energy Sciences, Scientific Discovery through Advanced Computing (SciDAC) program under Award Number DE-SC0022263. 
Work performed at the Center for Nanoscale Materials, a U.S. Department of Energy Office of Science User Facility, was supported by the U.S. DOE, Office of Basic Energy Sciences, under Contract No. DE-AC02-06CH11357. 

\newpage
\section*{Appendix A: Singular value decomposition}
For completeness, we outline the
approach taken for the construction of the inverse of the
complex (Hermitian) overlap matrix $\mathbf{S}$ based on singular value decomposition (SVD) as suggested in Ref. \onlinecite{klymko2021real}. The complex SVD of $\mathbf{S}$ is \cite{trefethen1997,businger1969}
\begin{equation}
    \mathbf{S} = \mathbf{U} \mathbf{D} \mathbf{V}^\dag  ,
\label{SVD}
\end{equation}
\noindent where in $\mathbf{U}$ and $\mathbf{V}$ are unitary matrices, $\mathbf{D}$ is a diagonal matrix with real elements (called the singular values), $\mathbf{D} = \rm{diag}(d_1,d_2,...,d_K)$, and $\dag$ denotes Hermitian conjugate (or conjugate transpose). In the present case, the dimension of all matrices is $K$ x $K$ where $K$ = $MR$ as discussed in the main text.  As the matrix $\mathbf{S}$ becomes ill-conditioned owing to linear dependencies in the Krylov vectors, one or more of the $\{d_j\}$ can become quite small relative to the largest singular values.  If a small positive threshold $\epsilon$ is introduced, one can set to zero all the $\{d_j\}$ in $\mathbf{D}$ that are $\leq \epsilon$ and still obtain an excellent description of $\mathbf{S}$ via Eq. \ref{SVD}.

The appropriate inverse of $\textbf{S}$ is then computed as
\begin{equation}
\mathbf{S}^{-1} = \mathbf{V} \mathbf{D}^{-1} \mathbf{U}^\dag   ,
\end{equation}
\noindent where $\mathbf{D}^{-1} = \rm{diag}(1/d_1,1/d_2 ...,1/d_K)$ ,
but such that whenever a $d_j$ is zero, the corresponding $1/d_j$ is
also set to zero.  This well-established approach in numerical analysis, while slightly non-intuitive, effectively represents $\mathbf{S}$ and $\mathbf{S}^{-1}$ in a lower dimension space that is not ill-conditioned. For the majority of the simulations in the numerical experiments section of the manuscript, a threshold of $\epsilon = 1e-9$ was used unless specified otherwise. 

\section*{Appendix B: Dipole moment operator}
The dipole moment coefficients, $\mu_{pq}$, consist of one-electron integrals which can be defined explicitly as, $\mu_{pq} = \int\! d\sigma \; \phi_p^*(\sigma)\left( -e \mathbf{r} \right) \phi_q(\sigma)$, where $\sigma$ is a generalized coordinate consisting of the spatial and spin degrees of freedom, $\sigma = (\mathbf{r},s)$, while the function $\phi(\sigma)$ represents a one-electron spin-orbital. These quantities can be calculated using standard Python packages such as PySCF\cite{Chan20_024109}. 

\section*{Appendix C: Relation to linear absorption spectrum}

Consider Fermi’s Golden Rule expression for the line shape function for incident radiation polarized in the $\xi$th direction ($\xi \in \{x, y, z\}$, $\hbar$ = 1):
\begin{equation}
    I_\xi(\omega) = \sum_{i,f} \rho_i |\braket{\psi_i|\hat{\mu}_\xi|\psi_f}|^2 \delta(E_f - E_i -\omega),
\end{equation}
where $E_i$ and $E_f$ correspond to the energies of the initial and final electronic states, and $\omega$ is the frequency of the incident radiation. Here, $\rho_i$ is the Boltzmann factor for describing a system initially in thermal equilibrium. At zero temperature, the Boltzmann factor is equal to one for the ground-state $\ket{\psi_G}$ and zero everywhere else, reducing the above equation to:
\begin{equation}
    I_\xi(\omega) = \sum_{f}  |\braket{\psi_G|\hat{\mu}_\xi|\psi_f}|^2 \delta(E_f - E_G -\omega).
\end{equation}
Using, $\delta(\omega) = \int_{-\infty}^\infty  e^{-i\omega t}\,dt $, we make the following simplifications:
\begin{align}
    I(\omega) &=   \int_{-\infty}^\infty  e^{-i(E_f - E_G - \omega) t}  \sum_{f}  |\braket{\psi_G|\hat{\mu}_\xi|\psi_f}|^2 \,dt \nonumber \\
    &=   \int_{-\infty}^\infty  e^{i(E_G + \omega) t}  \sum_{f}  \braket{\psi_G|\hat{\mu}_\xi e^{-i\hat{H}t}|\psi_f}\braket{\psi_f|\mu_\xi|\psi_G} \,dt \nonumber \\
    &=   \int_{-\infty}^\infty  e^{i(E_G + \omega) t}   \braket{\psi_G|\hat{\mu}_\xi e^{-i\hat{H}t}\mu_\xi|\psi_G} \,dt \nonumber \\    
    &= \int_{-\infty}^\infty  e^{i(E_G + \omega) t}  \braket{\mu_\xi(0)|\mu_\xi(t)} dt
\end{align}
where $\ket{\mu_\xi(t)} = e^{-i\hat{H}t}\hat{\mu}_\xi\ket{\psi_G}$. An absorption spectrum or oscillator strength can then be extracted from the real part of the lineshape function,
\begin{equation}
    f(\omega) = \frac{2}{3}\omega \sum_\xi \text{Re}[I_\xi(\omega)].
\end{equation}
This result shows how the dipole moment auto-correlation function is related to absorption spectrum. It should be noted that the initial state should ideally correspond to the electronic ground-state wavefunction.





%

\end{document}